\documentclass[9pt,conference]{IEEEtran}

\usepackage{waspaa25} %

\usepackage{mathtools}
\usepackage{bm} %
\usepackage{tikz}
\usetikzlibrary{fit,positioning,backgrounds,arrows.meta,calc,intersections,matrix}
\usepackage{pifont}
\usepackage{dsfont}
\usepackage{multirow}

\definecolor{indigo}{rgb}{0.29, 0.0, 0.51}
\definecolor{lavender}{rgb}{0.9, 0.9, 0.98}
\definecolor{mixTealOrange}{rgb}{0.9, 0.915, 0.85}

\def\promptcolor{teal!20}
\def\mixcolor{orange!20}
\def\combcolor{mixTealOrange}

\newcommand{\RR}{\mathbb{R}}
\newcommand{\tinytimes}{\mathbin{\mspace{-6mu}\times\mspace{-6mu}}}
\newcommand{\tinyadd}{\mathbin{\mspace{-6mu}+\mspace{-6mu}}}
\newcommand{\y}{\ding{51}}
\newcommand{\n}{\ding{55}}
 \newcommand{\miniscule}[1]{{\fontsize{3.5pt}{4.5pt}\selectfont #1}}

\DeclarePairedDelimiter\floor{\lfloor}{\rfloor}

\title{FasTUSS: Faster Task-Aware Unified Source Separation}

\name{Francesco Paissan$^{1,2}$,
      Gordon Wichern$^{1}$,
      Yoshiki Masuyama$^{1}$,
      Ryo Aihara$^{1}$,\\
      \emph{Fran\c{c}ois G.\ Germain$^{1}$,
      Kohei Saijo$^{3}$,
      Jonathan Le Roux$^{1}$}\thanks{This work was done while F.\ Paissan was an intern at MERL.}}
\address{$^{1}$Mitsubishi Electric Research Laboratories (MERL), Cambridge, MA, USA\\
$^{2}$University of Trento, Trento, Italy \quad $^{3}$Waseda University, Tokyo, Japan
}

\begin{document}
\bstctlcite{IEEEexample:BSTcontrol}
\maketitle

\begin{abstract}
    Time-Frequency (TF) dual-path models are currently among the best performing audio source separation network architectures, achieving state-of-the-art performance in speech enhancement, music source separation, and cinematic audio source separation. While they are characterized by a relatively low parameter count, they still require a considerable number of operations, implying a higher execution time. This problem is exacerbated by the trend towards bigger models trained on large amounts of data to solve more general tasks, such as the recently introduced task-aware unified source separation (TUSS) model. TUSS, which aims to solve audio source separation tasks using a single, conditional model, is built upon TF-Locoformer, a TF dual-path model combining convolution and attention layers.
    The task definition comes in the form of a sequence of prompts that specify the number and type of sources to be extracted. In this paper, we analyze the design choices of TUSS with the goal of optimizing its performance-complexity trade-off. We derive two more efficient models, FasTUSS-8.3G and FasTUSS-11.7G that reduce the original model's operations by 81\% and 73\% with minor performance drops of 1.2~dB and 0.4~dB averaged over all benchmarks, respectively. Additionally, we investigate the impact of prompt conditioning to derive a causal TUSS model.
\end{abstract}

\section{Introduction} \label{sec:intro}

With the advent of neural networks in audio source separation, different neural network architectures have been explored to solve specific tasks such as speech enhancement (SE) \cite{WDL2018,reddy2020interspeech,zhang2023toward}, speech separation (SS) \cite{dc,pit,dprnn,sepformer,tfgridnet}, music source separation (MSS) \cite{bsrnn,sawata2021all}, sound event separation \cite{universal_sound_separation,tzinis2020improving}, and cinematic audio source separation (CASS) \cite{zhang2021multi,petermann2022cocktail,Uhlich2024CDX}. 
There has recently been a shift towards solving general audio source separation (GASS) \cite{pons2024gass} by training a single model using data from multiple audio source separation tasks.
Based on the realization that GASS is an inherently ill-posed problem whose goal is task-dependent, task-aware unified source separation (TUSS) \cite{tuss} reformulates GASS as a conditional, task-aware, source separation problem, thereby solving multiple tasks using a single model. 
While conditional models have been used in the audio source separation literature for target sound extraction (TSE) \cite{vzmolikova2019speakerbeam,wang2022few,chen2022zero}, where the stems to be extracted can be specified by class IDs \cite{wang19h_interspeech,seetharaman2019class,ochiai20_interspeech,tzinis2022heterogeneous,delcroix2022soundbeam}, sound examples \cite{vzmolikova2019speakerbeam,chen2022zero,lee2019audio}, or even text queries \cite{kilgour22_interspeech,liu22w_interspeech,clapsep}, TSE models only extract one source, or a group of sources, at a time. Therefore, they do not model the relationship between queries, and thus are not task-aware. For example, speech sources should be treated differently when separating two overlapping voices and when separating all the speech in a podcast from the background music.

Formulating TUSS as a conditional source separation problem requires designing a model capable of (i) handling a variable number of output sources and (ii) adapting the source definition depending on the mixture and the input prompts. The model should also be powerful enough to handle a wide variety of task scenarios and to learn from large amounts of data. %
This made the TF-Locoformer architecture \cite{tflocoformer} an ideal building block for TUSS: encapsulating the modeling power of time-frequency (TF)-domain dual-path models for source separation \cite{bsrnn,tfgridnet,bsroformer} within a transformer-based architecture, it reaches state-of-the-art performance on multiple tasks while making it easy to introduce prompt tokens as a way to specify the task of interest.
Like all TF dual-path models, however, while TF-Locoformer has relatively few parameters (11.1M), it requires a considerable amount of operations, %
making model training more expensive
and incurring high computational cost at inference time.

In this paper, we explore ways to reduce the computational cost of the TUSS architecture, and to make it usable in practical real-time conditions.
Analyzing the performance of TUSS over five datasets and with different continuous source separation (CSS) settings (i.e., block size for separating long signals), we propose a set of optimizations for the TUSS architecture and benchmark their impact on the model's performance, leading to faster versions of the model, referred to as \emph{FasTUSS}. Additionally, we perform an extensive study to quantify the interplay between mixture and prompts in the TUSS architecture. We use the outcome of these experiments to derive a causal version of TUSS that can employ common optimizations such as KVCache~\cite{kvcache}.

\section{Overview of the TUSS Architecture}

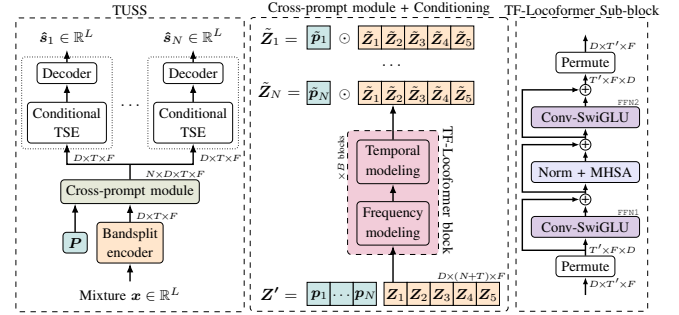
\begin{figure}
    \centering
    \resizebox{\linewidth}{!}{
        \begin{tikzpicture}
    \def\size{0.45}
    \def\S{3}
    \def\L{5}
    \def\blockwidth{40}
    \def\verticalskip{10pt}
    \def\hskiptse{35pt}

    \def\hcenter{2*0.5+0.5*\L*\size}

    \node (mix) at (-3, 0.5*\size) {\footnotesize Mixture $\bm{x}\in\RR^L$};

    \node[draw=black, fill=\mixcolor, rounded corners=2pt, above=\verticalskip of mix, align=center] (enc) {\footnotesize Bandsplit \\ \footnotesize encoder};
    \node[above=-2.5pt of enc, xshift=15pt] () {\tiny $D\tinytimes T\tinytimes F$};
    
    \node[draw=black, rounded corners=2pt, left=5pt of enc, fill=\promptcolor, align=center] (prompts) {\footnotesize $\bm{P}$};
    
    \node[draw=black, rounded corners=2pt, above=\verticalskip of enc, name path=xprompt_path, fill=\combcolor] (xprompt) {\footnotesize Cross-prompt module};
    \node[above=-2.5pt of xprompt, xshift=25pt] () {\tiny $N\tinytimes D\tinytimes T\tinytimes F$};
    
    \node[draw=black, rounded corners=2pt, above=\verticalskip+10pt of xprompt, align=center, xshift=-\hskiptse] (ctse1) {\footnotesize Conditional \\ \footnotesize TSE};
    
    \node[draw=black, rounded corners=2pt, above=\verticalskip of ctse1, align=center] (dec1) {\footnotesize Decoder};
    
    \node[rounded corners=2pt, above=0.7*\verticalskip of dec1, align=center] (s1) {\footnotesize $\bm{\hat{s}}_1\in\RR^L$};

    \node[right=5pt of ctse1, yshift=10pt, xshift=-0.5pt] {$\dots$};
    
    \node[draw=black, rounded corners=2pt, above=\verticalskip+10pt of xprompt, align=center, xshift=\hskiptse] (ctse2) {\footnotesize Conditional \\ \footnotesize TSE};
    
    \node[draw=black, rounded corners=2pt, above=\verticalskip of ctse2, align=center] (dec2) {\footnotesize Decoder};
    
    \node[rounded corners=2pt, above=0.7*\verticalskip of dec2, align=center] (s2) {\footnotesize $\bm{\hat{s}}_N\in\RR^L$};

    \node[draw, densely dotted, fit=(ctse1)(dec1), rounded corners=2pt] (fit1) {};
    \node[draw, densely dotted, fit=(ctse2)(dec2), rounded corners=2pt] (fit2) {};
    
    \node[below=0pt of fit1, xshift=15pt] () {\tiny $D\tinytimes T\tinytimes F$};
    \node[below=0pt of fit2, xshift=15pt] () {\tiny $D\tinytimes T\tinytimes F$};

    \draw[-{Latex[length=5pt]}, thick] (mix.north) -- (enc.south);
    \draw[-{Latex[length=5pt]}, thick] (enc.north) -- (xprompt.south);

    \coordinate (vprojection) at ($(prompts.north)+(0,4)$); 
    \draw[draw=none, name path=arrow] (prompts) -- (vprojection);
    \path [name intersections={of=arrow and xprompt_path, by={I1, I2}}];
    \draw[-{Latex[length=5pt]}, thick] (prompts.north) -- (I2);

    \node[above=2pt of xprompt] (ghst_xprompt) {};
    
    \draw[thick] (xprompt.north) -- (ghst_xprompt.north);
    \draw[-{Latex[length=5pt]}, thick] (ghst_xprompt.north) -| (ctse1.south);
    \draw[-{Latex[length=5pt]}, thick] (ghst_xprompt.north) -| (ctse2.south);

    \draw[-{Latex[length=5pt]}, thick] (ctse1.north) -- (dec1.south);
    \draw[-{Latex[length=5pt]}, thick] (dec1.north) -- (s1.south);
    \draw[-{Latex[length=5pt]}, thick] (ctse2.north) -- (dec2.south);
    \draw[-{Latex[length=5pt]}, thick] (dec2.north) -- (s2.south);
    
    \foreach \i in {1,...,\S} {
        \draw[fill=teal!20] (\i*\size, 0) rectangle ++(\size, \size);
        
        \ifnum\i=1
            \node at (0.02 + \i*\size + 0.5*\size, -0.02 + 0.5*\size) {\footnotesize $\bm{p}_{\i}$};
        \fi
        \ifnum\i=2
            \node at (0.02 + \i*\size + 0.5*\size, -0.02 + 0.5*\size) {\footnotesize $\dots$};
        \fi
        \ifnum\i=3
            \node at (0.02 + \i*\size + 0.5*\size, -0.02 + 0.5*\size) {\footnotesize $\bm{p}_{N}$};
        \fi
    }
    \foreach \i in {1,...,\L} {
        \draw[fill=orange!20] (1.5 + \i*\size, 0) rectangle ++(\size, \size);
        \node at (1.5 + \i*\size + 0.5*\size, -0.02+0.5*\size) {\footnotesize $\bm{Z}_{\i}$};
    }

    \node at (-0.1, 0.24) {$\bm{Z'}=$};
    \node at (1.4 + \L*\size, 0.55) {\tiny $D\tinytimes (N\tinyadd T)\tinytimes F$};
    
    \node at (\hcenter, \size-0.11) (ghst1) {};

    \node [rounded corners=2pt, minimum width=\blockwidth, align=center,  draw] (fmod) at (\hcenter, 1.6){\footnotesize Frequency \\ \footnotesize modeling};
    
    \node [rounded corners=2pt, minimum width=\blockwidth, align=center,  draw] (tmod) at (\hcenter, 2.8){\footnotesize Temporal \\ \footnotesize modeling};
    
    \begin{pgfonlayer}{background}
        \node [rounded corners=2pt, draw, dashed, fill=purple!20, fit=(tmod)(fmod), inner sep=5pt] (wrap) {};
    \end{pgfonlayer}
    \node[right=5pt of wrap, yshift=40pt, rotate=270] {\footnotesize TF-Locoformer block};
    
    \node [left=3pt of wrap, yshift=35pt, rotate=90] {\tiny $\times B$ blocks};

    \draw[fill=teal!20] (\size, 1.4*2.8) rectangle ++(\size, \size);
    \node at (0.02 + \size + 0.5*\size, -0.02 + 0.5*\size + 1.4*2.8) {\footnotesize $\tilde{\bm{p}}_{N}$};
    \node at (0.06 + 2*\size + 0.5*\size, -0.02 + 0.5*\size + 1.4*2.8) {\footnotesize $\odot$};
    
    \foreach \i in {1,...,\L} {
        \draw[fill=orange!20] (1 + \i*\size, 1.4*2.8) rectangle ++(\size, \size);
        \node at (1.02 + \i*\size + 0.5*\size, -0.02+0.5*\size+1.4*2.8) {\footnotesize $\tilde{\bm{Z}}_{\i}$};
    }

    \node at (\hcenter, 1.4*2.8+0.11) (ghst2) {};
    \node at (-0.1, 1.4+2.8) {$\tilde{\bm{Z}}_N=$};

    \draw[fill=teal!20] (\size, 5) rectangle ++(\size, \size);
    \node at (0.02 + \size + 0.5*\size, 0.5*\size + 5) {\footnotesize $\tilde{\bm{p}}_{1}$};
    \node at (0.06 + 2*\size + 0.5*\size, -0.02 + 0.5*\size + 5) {\footnotesize $\odot$};
    
    \foreach \i in {1,...,\L} {
        \draw[fill=orange!20] (1 + \i*\size, 5) rectangle ++(\size, \size);
        \node at (1.02 + \i*\size + 0.5*\size, -0.02+0.5*\size+5) {\footnotesize $\tilde{\bm{Z}}_{\i}$};
    }

    \node at (-0.1, 5 + 0.5*\size+0.01) {$\tilde{\bm{Z}}_1=$};
    
    \node at (\hcenter, 4.45 + 0.5*\size+0.01) {$\dots$};

    \draw[-{Latex[length=5pt]}, thick] (ghst1) -- (fmod.south);
    \draw[-{Latex[length=5pt]}, thick] (fmod.north) -- (tmod.south);
    \draw[-{Latex[length=5pt]}, thick] (tmod.north) -- (ghst2);

    \node (zprime) at (\hcenter + 3.75, 0.2*\size) {};

    \node[draw=black, rounded corners=2pt, above=\verticalskip of zprime] (perm) {\footnotesize Permute};
    \node[above=-2.5pt of perm, xshift=15pt] () {\tiny $T'\tinytimes F\tinytimes D$};
    \node[below=-2.5pt of perm, xshift=15pt] () {\tiny $D\tinytimes T'\tinytimes F$};
    
    \node[draw=black, rounded corners=2pt, above=\verticalskip of perm, minimum width=60pt, fill=indigo!20] (ffn1) {\footnotesize Conv-SwiGLU};
    \node[draw=black, circle, above=0.6*\verticalskip of ffn1, inner sep=0pt,] (p1) {\footnotesize +};
    \node[above=-2.5pt of ffn1, xshift=24pt] {\tiny \texttt{FFN1}};

    \node[draw=black, rounded corners=2pt, above=0.6*\verticalskip of p1, minimum width=60pt, fill=lavender] (mhsa) {\footnotesize Norm + MHSA};
    \node[draw=black, circle, above=0.6*\verticalskip of mhsa, inner sep=0pt,] (p2) {\footnotesize +};

    \node[draw=black, rounded corners=2pt, above=0.6*\verticalskip of p2, minimum width=60pt, fill=indigo!20] (ffn2) {\footnotesize Conv-SwiGLU};
    \node[draw=black, circle, above=0.6*\verticalskip of ffn2, inner sep=0pt,] (p3) {\footnotesize +};
    \node[above=-2.5pt of ffn2, xshift=24pt] {\tiny \texttt{FFN2}};
    
    \node[draw=black, rounded corners=2pt, above=0.6*\verticalskip of p3] (perm2) {\footnotesize Permute};
    \node[above=-2.5pt of perm2, xshift=15pt] () {\tiny $D\tinytimes T'\tinytimes F$};
    \node[below=-2.5pt of perm2, xshift=15pt] () {\tiny $T'\tinytimes F\tinytimes D$};
        
    \node[above=\verticalskip of perm2] (zprime2) {};

    \draw[-{Latex[length=4pt]}, thick] (zprime.north) -- (perm.south);
    \draw[-{Latex[length=4pt]}, thick] (perm.north) -- (ffn1.south) coordinate[pos=0.3] (m1);
    \draw[-{Latex[length=4pt]}, thick] (ffn1.north) -- (p1.south);
    \draw[-{Latex[length=4pt]}, thick] (m1) -- ++(-35pt, 0) |- (p1.west);
    \draw[-{Latex[length=4pt]}, thick] (p1.north) -- (mhsa.south) coordinate[pos=0.15] (m2);
    \draw[-{Latex[length=4pt]}, thick] (m2) -- ++(-35pt, 0) |- (p2.west);
    \draw[-{Latex[length=4pt]}, thick] (mhsa.north) -- (p2.south);
    \draw[-{Latex[length=4pt]}, thick] (p2.north) -- (ffn2.south) coordinate[pos=0.15] (m3);
    \draw[-{Latex[length=4pt]}, thick] (m3) -- ++(-35pt, 0) |- (p3.west);
    \draw[-{Latex[length=4pt]}, thick] (ffn2.north) -- (p3.south);
    \draw[-{Latex[length=4pt]}, thick] (p3.north) -- (perm2.south);
    \draw[-{Latex[length=4pt]}, thick] (perm2.north) -- (zprime2.south);

    \node[draw, fit=(mix)(s1)(s2)(fit1)(fit2), rounded corners=2pt, dashed, name path=APath] (A) {};
    \node[above=-2pt of A] {\footnotesize TUSS};

    \node[minimum width=143pt, minimum height=163pt, draw, dashed, rounded corners=2pt] at (\hcenter - 0.27, 2.75) (B) {};
    \node[above=-2pt of B, align=center] {\footnotesize Cross-prompt module + \footnotesize Conditioning};
    
    \node[minimum width=73pt, minimum height=163pt, draw, dashed, rounded corners=2pt] at (\hcenter + 3.68, 2.75) (C) {};
    \node[above=-2pt of C, align=center] {\footnotesize TF-Locoformer \footnotesize Sub-block};

\end{tikzpicture}
    }
    \vspace{-15pt}
    \caption{TUSS architecture (left panel). Cross-prompt module and conditioning strategy for $T=5, N=5$ (middle panel). Each TF-Locoformer block has frequency modeling and temporal modeling sub-blocks with the same architecture (right panel), where $T'=N + T$ for the cross-prompt module and $T'=T$ inside conditional TSE.}
    \label{fig:tuss}
    \vspace{-9pt}
\end{figure}

The TUSS architecture, depicted in \cref{fig:tuss}, takes as input a mixture $\bm{x}\in\RR^{L}$ of length $L$ and a set of $N$ prompts indicating the set of target sources (or stems) that we want to extract from the mixture. For example, the network can separate a scene into stems consisting of all speakers, all music, and all sound effects when prompted with \texttt{[Speech, Music-mix, SFX-mix]}, 
but can further separate the instruments given the prompts \texttt{[Speech, Drums, Vocals, Bass, Other inst., SFX-mix]}. 
The mixture $\bm{x}$ is first processed using a short-time Fourier transform (STFT) and a band-split encoder as in \cite{bsrnn,bsroformer}, yielding $\bm{Z}\in\RR^{D\times T\times F}$, where $D, T, F$ represent the number of channels, frames, and frequency bands, respectively. The encoded mixture $\bm{Z}$ is then processed, together with a set of $N$ learned prompt embeddings $\bm{P}=[\bm{p}_1,\dots,\bm{p}_N] \in \RR^{D\times N}$ corresponding to the input prompts, in two stages: a cross-prompt module followed by a conditional TSE module. 
In the cross-prompt module, the set of prompts is prepended to the encoded mixture on the frame axis, yielding $\bm{Z'}=[\bm{P}\;\bm{Z}]\in\RR^{D\times (N+T)\times F}$ after appropriate broadcasting, and
$\bm{Z'}$ is input to Transformer-based blocks to model the dependency of the temporal sequence.
The conditioning between prompts and mixture happens bidirectionally in the multi-head self-attention blocks. Injecting prompts in this way enables conditioning based on an arbitrary number of sources and specializing each prompt for the set of prompts and the mixture currently being processed.

The conditional TSE module then extracts the source specified by each prompt in parallel. 
The output $\bm{\tilde{Z}'} = [\tilde{\bm{P}}\;\tilde{\bm{Z}}]$ of the cross-prompt module is split into the encodings of the prompts and the mixture, which are combined using element-wise product. This results in a mixture representation conditioned by each prompt, $\tilde{\bm{Z}}_n=\tilde{\bm{Z}}\odot\tilde{\bm{p}}_n$, depicted in the middle panel of \cref{fig:tuss}. 
Each $\tilde{\bm{Z}}_n$ is further processed by a sequence of Transformer blocks, similarly to the cross-prompt module. The output of the conditional TSE module is fed to a decoder that processes the sequence via an MLP and inverse STFT, resulting in a separated signal $\hat{\bm{s}}_n\in\RR^{L}$ for each prompt. 
The same conditional TSE module and decoder are used for all prompts.

In both stages, the Transformer-based processing consists in stacks of TF-Locoformer blocks. As illustrated in the middle panel of \cref{fig:tuss} for the case of the cross-prompt module, within each TF-Locoformer block, the input is processed frame-wise in a frequency modeling path then frequency-wise in a temporal modeling path, where both the frequency modeling and the temporal modeling consist of TF-Locoformer sub-blocks, depicted in the right part of \cref{fig:tuss}.  Within the cross-prompt module, the TF-Locoformer sub-blocks on the temporal modeling path process the mixture using pointwise convolutions (in both \texttt{FFN1} and \texttt{FFN2} in the right part of \cref{fig:tuss}) to make the network robust to the order of the prompts in $\bm{P}$. 

\begin{figure}
    \centering
    \includegraphics[width=.96\linewidth]{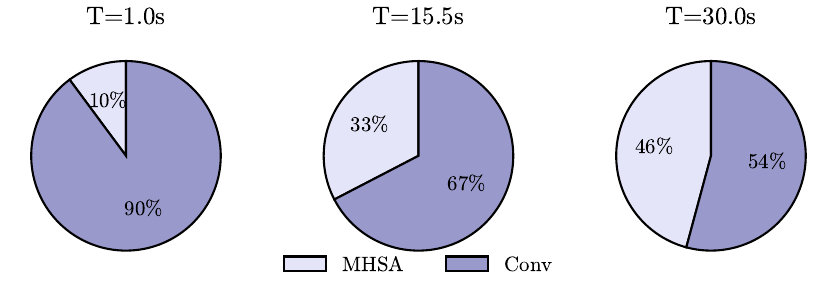}
    \vspace{-.3cm}
    \caption{Compute breakdown of MHSA vs.\ convolutions for different chunk lengths.}
    \vspace{-.3cm}
    \label{fig:profile}
\end{figure}

\section{Profiling and optimizing TUSS} \label{sec:profile}
To derive a faster version of TUSS, we start by analyzing the computational bottleneck of the original model. We measure the number of multiply-accumulate (MAC)\footnote{We use MAC as an indirect measure of inference time.} operations in TUSS and analyze the percentage of compute required by convolutions vs.\ multi-head self-attention (MHSA) within the TF-Locoformer block. The results of this profiling, summarized in \cref{fig:profile}, showcase that the percentage varies based on sequence length, because the computation required to perform a convolution scales linearly with sequence length, while it scales quadratically for MHSA. Notably, we find that, especially for short audio chunks, the majority of the compute is spent for convolutions (90\% for \SI{1}{\second} of audio), while the contributions of MHSA and convolutions become more comparable for sequences of \SI{30}{\second}.
Guided by this finding, we focus on optimizing the convolutional part of the TF-Locoformer block, depicted in the right part of \cref{fig:tuss}.

The convolutions inside the TF-Locoformer block are within the Conv-SwiGLU blocks. Each of these blocks comprises a normalization layer, one convolution with Swish activation, another convolution used for gating, and a deconvolution. In the original TUSS paper~\cite{tuss}, all convolutions except those processing the frame path in the cross-prompt module use one-dimensional kernels of size 4, and a hop size of 1. The deconvolution uses the same parameters to ensure the output dimensionality matches the input.
Recall that the computational cost, in MAC, of a one-dimensional grouped convolution $\Psi: \RR^{C_\text{in}\times L}\to\RR^{C_\text{out}\times L'}$ is
\begin{align} \label{eq:comp}
    L' \frac{C_\text{in}C_\text{out} K}{G} = \floor*{\frac{L + 2p - K}{S} + 1} \frac{C_\text{in}C_\text{out}K}{G},
\end{align}
where $G$ is the number of groups, $S$ is the stride, $p$ is the amount of padding applied, and $K$ is the kernel size.
We leave $p=\floor{(K-1)/2},\,K=4$ to avoid deviating too much from the original TUSS architecture, and observe that all hyper-parameters linearly scale the complexity of the convolution.
Therefore, to optimize the overall structure of the FFN blocks, we investigate the impact of changing (a)~the stride: we test configurations with $S\in\{1,2,4\}$; (b)~the number of groups: we analyze the impact of grouped convolutions with $G=8$ and channel shuffle, to ensure features are distributed from one group to the others \cite{shufflenet,shufflenetv2,Li2021MicroNetII}, and we test the performance of depthwise separable convolutions ($G=C_\text{in}$) followed by a pointwise convolution ($K=1$), a commonly used configuration for resource-efficient neural networks \cite{Chollet2016XceptionDL,Lin2020MCUNetTD,Paissan2021PhiNetsAS,tzinis2020}; and (c)~the use of two FFN blocks before and after MHSA: we remove the first or second FFN block and analyze the performance of combinations of these optimizations. %

We also explore variants of MHSA similar to \cite{subakan2023} but with more recent techniques.
We tested replacing the MHSA in the TF-Locoformer block with (a) linear unified nested attention (LUNA) \cite{Ma2021LunaLU}, (b) cascaded grouped attention (CGA) \cite{Liu2023EfficientViTME}, and (c) a bi-directional GRU. We emphasize that these optimizations result in minimal improvements in terms of compute, especially for short audio segments. Nonetheless, they could be beneficial for reducing the working memory requirement of the architecture. We observe that reducing the capacity of the TF-Locoformer block by replacing the MHSA affects its sequence modeling capabilities. This is most relevant within the temporal modeling path of the cross-prompt module, which also lacks the convolutional FFNs. To alleviate this, we propose (d) a prompt-aware Conv-SwiGLU block: this block is a modified version of the original FFN block, in which %
we process the prepended prompts via linear layers, while the mixture is processed using the original Conv-SwiGLU structure, with $K=4$. 

Results are discussed in \cref{sec:compression}.

\section{CAUSAL TUSS}

To derive a causal version of TUSS, we make all the blocks within the architecture causal. While the convolutional part is straightforward, the MHSA inside the cross-prompt module needs special care to ensure the mixture processing is causal and the prompts are updated without leaking information during the training stage. For example, we note that using a standard attention mask on the prompts would lead to the first prompt (first element of $\bm{Z}'$) never being updated with information from other prompts or the mixture.
Recall that the attention map of the $h$-th head in MHSA is computed as
\begin{align}
    \bm{\Lambda}_h = \text{Softmax}\left(\frac{\bm{Q}_h \bm{K}_h^\top}{\sqrt{D}}\right) \in \RR^{(N+T)\times (N+T)},
\end{align}
where $\bm{Q}_h = \bm{X} \bm{W}_h^Q $ and $\bm{K}_h=\bm{X} \bm{W}_h^K \in \RR^{(N+T)\times D}$ are learnable linear projections of the MHSA layer's input $\bm{X}\in\RR^{(N+T)\times C_\text{in}}$, and Softmax is applied row-wise.
We want to derive an attention mask $\bm{M}\in\RR^{(N+T)\times (N+T)}$ to modify the attention map via element-wise multiplication,
such that $M_{ij} = 1$ if token $j$ influences the update of token $i$, $M_{ij}=0$ otherwise\footnote{We will informally say that token $i$ `sees' token $j$ when $M_{ij}=1$.}. Specifically, we want $\bm{M}$ to (i) mimic causal inference, (ii) minimally impact the model's performance, and (iii) enable inference using KVCache \cite{kvcache}. To achieve the objectives above, we investigate the impact of the prompts in the mixture's processing and vice versa by experimenting with different attention mask designs. %

\begin{figure}
    \centering
    \resizebox{0.97\linewidth}{!}{
        \input{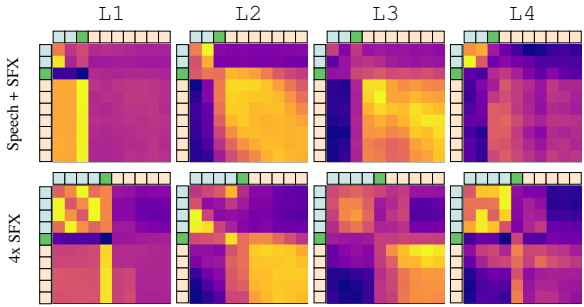}
    }
    \vspace{-0.3cm}
    \caption{
        Attention maps in log scale. Teal, green, and orange tokens represent the prompts, the \texttt{<SOS>} token, and the mixture, respectively. %
        \texttt{L*} represent the layers %
        on the temporal modeling path of the cross-prompt module.
    }
    \label{fig:attn_viz}
    \vspace{-9pt}
\end{figure}

\begin{figure}
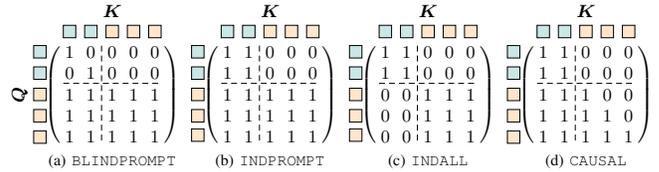

    \centering
    \resizebox{\linewidth}{!}{
        \begin{tikzpicture}
    \def\L{2}
    \def\size{0.7em}

    \input{figs/maps/blind}
    \input{figs/maps/indprompt}
    \input{figs/maps/indall}
    \input{figs/maps/causal}

\end{tikzpicture}
    }
    \vspace{-20pt}
    \caption{
        Attention masks used to investigate the impact of prompt-mixture conditioning and to train the causal model.
        Color coding follows \cref{fig:tuss}. We omit the \texttt{<SOS>} token from the visualization for brevity.
    } %
    \label{fig:attention_mask}
    \vspace{-9pt}
\end{figure}

We construct the attention masks for the experiments by splitting them into four blocks:
\begin{align}
    \bm{M} =
    \left(\begin{matrix}
        \bm{A} & \bm{B} \\
        \bm{C} & \bm{D} \\
    \end{matrix}\right),
\end{align}
where $\bm{A} \in\RR^{N\times N}$, $\bm{B}\in\RR^{N\times T}$, $\bm{C}\in\RR^{T\times N}$, and $\bm{D}\in\RR^{T\times T}$.
$\bm{A}$ controls which prompts are seen during each prompt's update. $\bm{B}$ controls which portions of the mixture influence the prompts. Finally, $\bm{C}$ and $\bm{D}$ represent the influence of the prompts and the mixture on the mixture's processing, respectively.
Based on a qualitative analysis of the attention masks in TUSS, depicted in \cref{fig:attn_viz}, we experiment with three TUSS variants. Specifically, prompt and mixture updates seem loosely correlated (i.e., the matrices look roughly block-diagonal). Therefore, we train the following models: 
\texttt{BLINDPROMPT}, where the prompts do not see each other, 
with $A_{ij}=\delta_{ij}$, $B_{ij}=0$, $C_{ij}=D_{ij}=1$, 
illustrated in \cref{fig:attention_mask}(a); 
\texttt{INDPROMPT}, where the prompts cannot see the mixture (i.e., the prompts are independently processed), 
with $B_{ij}=0$ and $A_{ij}=C_{ij}=D_{ij}=1$, shown in \cref{fig:attention_mask}(b); 
and \texttt{INDALL}, where the prompts cannot see the mixture and vice versa (i.e., they are independently processed from each other), 
with $A_{ij}=D_{ij}=1$, $B_{ij}=C_{ij}=0$, 
shown in \cref{fig:attention_mask}(c).

Finally, we derive a causal version of TUSS (\texttt{CAUSAL}) that can work with KVCache by making the prompts independent of the mixture, and making the mixture processing causal, as showcased in \cref{fig:attention_mask}(d). Formally, the last configuration is $B_{ij}=0$, $A_{ij}=C_{ij}=1$, $D_{ij}=\mathds{1}_{i\ge j}$. Using this attention mask is equivalent to performing inference of the MHSA over the entire sequence of prompts, then, using the same MHSA weights, process the mixture using KVCache by feeding one frame at a time, with the first one being appended to the prompts. This ensures that the prompts are attended to also during mixture processing, although without being updated.

Results are discussed in \cref{sec:causality_results}.

\section{Experimental setup}

\noindent \textbf{Datasets.} We follow the experimental setup presented in TUSS \cite{tuss}. During training, we create mixtures on the fly using samples from VCTK \cite{veaux2013voice},  WSJ0 \cite{wsj0}, LibriVox (from the URGENT challenge) \cite{zhang2024urgent}, FSD50k \cite{fonseca2021fsd50k}, WHAM! \cite{wichern19_interspeech}, DEMAND \cite{thiemann2013diverse}, MUSDB-HQ \cite{MUSDB18HQ}, MOISESDB \cite{pereira2023moisesdb}, and FMA \cite{defferrard2016fma}. 
We use eight prompt categories, \texttt{Speech}, \texttt{SFX}, \texttt{SFX-mix}, \texttt{Bass}, \texttt{Drums}, \texttt{Vocals}, \texttt{Other inst.}, and \texttt{Music-mix}. We sample the number of sources, $N$, for each training sequence from a multinomial distribution with $P(N=i)=1/3, \,\forall i \in \{2,3,4\}$. The only stems that can be repeated in a mixture are \texttt{Speech} and \texttt{SFX}.
For validation, we benchmark on multiple separation tasks using five datasets: VCTK-DEMAND for SE, WHAM! for noisy SS, FUSS for sound event separation, MUSDB-HQ for MSS, and DnR for CASS. We used the average SNR on the validation datasets to analyze the performance-complexity trade-offs of different architectures.
We invite the reader to refer to the original paper \cite{tuss} for more details on the dynamic mixing and the pre-processing pipelines.

\noindent \textbf{Hyperparameters.} We train all models for 375k steps. We employ the AdamW optimizer with weight decay set to $0.01$. We warm up the learning rate from $0$ to $0.001$ over 10k steps. After that, we decrease the learning rate using a cosine annealing learning rate schedule with $\eta_\text{min}=\num{5e-5}$. We found that the cosine schedule yields more consistent results among different model scales. We clip the gradients with a maximum $L_2$ norm of 5. We use a batch size of 4 for all models and scale the number of GPUs for the larger model experiments. We use the signal-to-noise ratio as training objective. During evaluation, we average the weights of the ten best checkpoints. 

\noindent \textbf{Models.} We use a band-split encoder and band-wise decoding module \cite{bsrnn} with $F=61$ bands to efficiently handle data with high sampling rate, as in \cite{tuss, bsroformer}. We tested the model optimization strategies described in \cref{sec:profile} on the TUSS \texttt{M5} configuration. Following the TF-Locoformer paper's notation \cite{tflocoformer}, \texttt{M5} uses $B = 4$, $D = 64$, $C = 384$, $K = 4$, $S = 1$, $H = 4$, $G = 8$, and attention head size $E=256$ in the cross-prompt module. The conditional TSE blocks, instead, use $B = 2$, $C = 256$, and $E = 96$.
To make the experimental setup more complete, we tested TUSS with and without a start-of-sentence \texttt{<SOS>} token applied between the prompts and the sequence in $\bm{Z}'$. We observed a performance decrease without the \texttt{<SOS>} token and thus use it in our experiments. Following \cite{saijo2025comparative}, we also investigate the model's performance without rotary positional encoding (RoPE) \cite{Su2021RoFormerET} and observe that positional encoding is beneficial in our setting. For LUNA, we use a hidden sequence length that scales proportionally to the input sequence length ($0.25\times$ the input length). For grouped self-attention, we use eight groups. In the GRU tests, we use a 1-layer, bi-directional GRU with the hidden dimension matching the attention dimension of the MHSA it replaces. 

\section{Results}
\label{sec:results}

\begin{table}[t]
\sisetup{
detect-weight,
mode=text,
tight-spacing=true,
round-mode=places,
round-precision=1,
table-format=2.1,
table-number-alignment=center
}
\centering
\caption{
    Comparison of various speedup and conditioning configurations.
    $^\dag$~indicates additional use of pointwise convolutions.
    MAC reported for \SI{1}{\second} of audio. $^\P$ refers to the configuration with prompt-aware FFN.
}
\resizebox{\linewidth}{!}{
    \begin{tabular}{ccccSSS[table-format=1.1]}
    \toprule
    
    \texttt{ID} & $S$ & $G$ & \!\texttt{FFN1}\! & \!{Params (M)}\! & \!{MAC (G)}\! & \!{$\Delta$ SNR [$\si{\deci\bel}$]}\! \\ \midrule
    \texttt{1} & 1 & 1 & \y & 11.1 & 43.1 & 0 \\
    \texttt{2} & 2 & 1 & \y & 11.1 & 26.2 & -0.2 \\
    \texttt{3} & 4 & 1 & \y & 11.1 & 17.7 & -0.5 \\
    \texttt{4} & 1 & 8 & \y & 10.8 & 40.5 & -1.6 \\
    \texttt{5} & 1 & 1 & \n & 8.9 & 24.4 & -0.3 \\
    \texttt{6} & 2 & 1 & \n & 8.9 & 16.0 & -0.6 \\
    \texttt{7} & 4 & 1 & \n & 8.9 & 11.7 & -0.4 \\
    \texttt{7}$^\P$\hspace*{-0.64em} & 4 & 1 & \n & 9 & 11.7 & -0.3 \\
    \texttt{8} & 4 & 8 & \n & 7.5 & 8.3 & -1.2 \\
    \texttt{9} & 4 & $C_\text{in}^\dag$ & \n & 7.4 & 8.6 & -1.2 \\
    
    \midrule
    
    \texttt{BLINDPROMPT} & 1 & 1 & \y & 11.1 & 43.1 & -1.3 \\
    \texttt{INDPROMPT} & 1 & 1 & \y & 11.1 & 43.1 & -0.2 \\
    \texttt{INDALL} & 1 & 1 & \y & 11.1 & 43.1 & -3.2 \\
    \texttt{CAUSAL} & 1 & 1 & \y & 11.1 & 43.1 & -1.8 \\
    
    \bottomrule
\end{tabular}

}
\label{tab:checkmarked-res}
\vspace{-11pt}
\end{table}

\subsection{Model speedup}
\label{sec:compression}

To evaluate different model optimizations, we consider their impact on test set SNR averaged over all benchmarks. We report the performance drop in \cref{tab:checkmarked-res} and refer to the different configurations' IDs during the presentation of the results\footnote{For the interested reader, we report the per-benchmark results in the Supplementary Material.}. \texttt{ID1} corresponds to the original \texttt{M5} TUSS model.
We observe that, on \SI{1}{\second} of audio, changing the stride (\texttt{ID2}, \texttt{ID3}) linearly reduces the number of operations required for the model, as expected from observing that the convolutions account for most of the compute (\cref{fig:profile}) and from \cref{eq:comp}. This optimization does not harm performance significantly, with the maximum stride configuration ($K=S=4$) reducing performance by \SI{0.5}{\dB}.
A comparable compute benefit comes from removing the first Conv-SwiGLU block, \texttt{FFN1}, from the original configuration. This nearly halves the number of operations, with a minimal SNR drop (\SI{0.3}{\dB}, \texttt{ID5}).
We analyzed the compound effect of these two optimizations by testing the model without \texttt{FFN1} for all stride configurations (\texttt{ID6}, \texttt{ID7}). Again, the impact of the stride is to reduce operations linearly. Even in this setting, we observe that increasing the stride results in a minimal performance decrease. Notably, the model without \texttt{FFN1} and with maximum stride (\texttt{ID7}) has about $25\%$ of the MAC of the original \texttt{M5} configuration, with a minimal performance reduction of \SI{0.4}{\dB}. For completeness, even though it is not standard in the literature, we tested the configuration where we remove \texttt{FFN2} instead, which resulted in worse SNR values.
Despite the performance drop that we observe on the \texttt{M5} configuration with groups and channel shuffle (\SI{1.6}{\dB}, \texttt{ID4}), we tested the performance of groups and channel shuffle on the model with maximum stride and without \texttt{FFN1}. We observe that the drop is lower in this setting (\SI{1.2}{\dB}, \texttt{ID8}), possibly because of the specific training hyperparameters that we used. We tested the model with depthwise-separable (DWS) convolutions (\texttt{ID9}), and observed a slightly higher number of MAC due to the introduction of the pointwise convolutions, with a similar drop. 
Given this analysis, we define \emph{FasTUSS-11.7G} as configuration \texttt{ID7} and \emph{FasTUSS-8.3G} as configuration \texttt{ID8} from \cref{tab:checkmarked-res}, as both are good options for maintaining strong performance while reducing compute.

We tested the impact of the prompt-aware FFN block on \texttt{ID7}. Despite a marginal increase in compute, performance was slightly improved (\SI{0.3}{\dB} below \texttt{M5}). This likely depends on the local modeling performed by the convolutions before the MHSA block, similarly to what was observed in \cite{tflocoformer}. However, we note that this configuration has more parameters due to the additional linear layers used to process the prompts independently of the mixture. We used this configuration to train model variants with LUNA and CGA, which were both outperformed by MHSA. Notably, 
both of these attention variants did not even provide significant computational benefits due to the small percentage of MAC allocated to attention and the relatively short sequence lengths.

To validate the performance of TUSS in continuous source separation settings (i.e., when the audio file is chunked and processed using a sliding window), we used the MUSDB-HQ and DnR benchmarks and report the average SNR values in \cref{tab:css}. We observe that TUSS consistently performs better with smaller window shifts (i.e., higher overlap), likely due to the correction of border effects using overlap-add. To validate the choice of optimizing the architecture for shorter chunk sizes, where more compute goes into convolutions than into attention (i.e., chunk size less than \SI{30}{\second}), we test with varying chunk size between \SI{4}{\second} and \SI{12}{\second} and observe that the performance plateaus after \SI{6}{\second}. This is likely linked to the chunk size used during training or the use of RoPE in TUSS \cite{saijo2025comparative}. 

\begin{table}[t]
    \centering
    \caption{
        Comparison with \texttt{ID1} with different CSS settings on MUSDB-HQ and DnR. MAC are reported for \SI{60}{\second} audio sequences.
    }
    \begin{tabular}{SSSS}
        \toprule
        {Chunk length [\si{\second}]} & {Overlap [\%]} & {MAC (T)} & {SNR [\si{\dB}]} \\
        \midrule
        4  & 0  & 2.7  & 7.5 \\
        4  & 50 & 5.3  & 7.8 \\
        6  & 0  & 2.8  & 7.7 \\
        6  & 50 & 5.4  & \textbf{8.0} \\
        6  & 75 & 10.5 & \textbf{8.0} \\
        8  & 0  & 2.8  & 7.7 \\
        10 & 0 & 3.1  & 7.6 \\
        12 & 0 & 3.2  & 7.3 \\
        \bottomrule
    \end{tabular}
    \label{tab:css}
    \vspace{-7pt}
\end{table}

\subsection{Causality and Conditioning Ablation}
\label{sec:causality_results} 

\Cref{tab:checkmarked-res} also shows the results for the models trained with various attention masks.
The \texttt{BLINDPROMPT} model, where the prompts are unable to see each other, shows a performance drop of \SI{1.3}{\dB}, suggesting that the prompts do benefit from looking at each other. 
This is somewhat expected, especially in the setting with repeated prompts (e.g., second row of \cref{fig:attn_viz}), where the prompts benefit from attending to each other to indicate different sources of the same type. 
For the \texttt{INDPROMPT} model, 
where the prompts never look at the mixture for their updates 
we only observe a minimal performance drop of \SI{0.3}{\dB}. 
However, the \texttt{INDALL} model, where the mixture also never looks at the prompt, does suffer from a significant performance drop, with average SNR decreasing by \SI{3.2}{\dB}.
This suggests that, while the conditioning of the mixture based on the prompt is critical, the model is not benefiting much from the conditioning of the prompts based on the mixture.
We can thus envision a causal model where the prompts never look at the mixture but the mixture looks at the prompt and is updated causally. For this \texttt{CAUSAL} model, 
we observe a total SNR drop of \SI{1.8}{\dB}.

The minimal drop in performance observed when the prompt sequence is processed independently of the mixture motivates the exploration of more resource-efficient sequence models for the mixture processing.
We consider replacing the MHSA block in the \texttt{ID7} configuration with a `hybrid' block, where MHSA processes the prompts while the mixture (with the prepended prompts) is processed using a bi-directional GRU. 
We observe however that MHSA is key to TUSS's performance, with performance dropping by \SI{0.8}{\dB} from the \texttt{ID7} model when replacing MHSA with GRU for mixture processing only in the hybrid model, and by \SI{1.4}{\dB} when MHSA is also replaced with GRU for prompt modeling.

\section{Conclusion}

In this paper, we presented FasTUSS, faster variants of the TUSS model for task-aware unified source separation. To derive the FasTUSS configurations, we benchmarked several optimizations of the TUSS architecture on five source separation tasks. Additionally, we presented the results of an ablation targeted at quantifying the interplay between the prompt and the mixture within the cross-prompt module. Using the results of the ablation, we derived a causal version of TUSS that can be implemented using modern inference optimizations such as KVCache.
Future work includes exploring device-specific optimization for real-time inference on edge devices. %

\clearpage
\bibliographystyle{IEEEtran}
\bibliography{refs25}

\clearpage
\onecolumn
\appendices
 
\section{Performance breakdown for all benchmarks}

\begin{table*}[h!]
\centering
\sisetup{
detect-weight, %
mode=text, %
tight-spacing=true,
round-mode=places,
round-precision=1,
table-format=2.1,
table-number-alignment=center
}
\caption{
    Evaluation results of the model configurations in \cref{tab:checkmarked-res}. SNR [dB] is reported for all datasets. We report MAC for \SI{1}{\second} except when differently specified.
}
\label{table:results}
\resizebox{\linewidth}{!}{
\begin{tabular}{cS[table-format=4.1]SSSSSS[table-format=1.1]SS*{4}{S[table-format=1.1]}S*{2}{S[table-format=1.1]}}
    \toprule
    
    & & & & & \multicolumn{2}{c}{VCTK-DEMAND (SE)} & \multicolumn{2}{c}{WHAM! (SS)} & {FUSS} & \multicolumn{4}{c}{MUSDB-HQ (MSS)} & \multicolumn{3}{c}{DnR (CASS)} \\
    \cmidrule(lr){6-7} \cmidrule(lr){8-9} \cmidrule(lr){10-10} \cmidrule(lr){11-14} \cmidrule(lr){15-17}
    
    & {MAC (G)} & {Params (M)} & {Chunk} & {Shift} & {Speech} & {SFX-mix} & {Speech} & {SFX-mix} & {SFX} & {Vocals} & {Bass} & {Drums} & {Other} & {Speech} & {Music-mix} & {SFX-mix} \\
    
    \midrule
    
    \texttt{ID1} & 43.1 & 11.1 & \SI{6}{\second} & \SI{6}{\second} & 19.6 & 10.9 & 7.7 & 11.5 & 12.4 & 7.7 & 5.4 & 7.9 & 4.5 & 14.4 & 6.5 & 7.5 \\
    \texttt{ID2} & 26.2 & 11.1 & \SI{6}{\second} & \SI{6}{\second} & 19.3 & 10.9 & 7.5 & 11.2 & 12.3 & 7.4 & 5.4 & 7.7 & 4.3 & 14 & 6.4 & 7.5 \\
    \texttt{ID3} & 17.7 & 11.1 & \SI{6}{\second} & \SI{6}{\second} &  19.1 & 10 & 6.7 & 10.9 & 12.4 & 7.1 & 5.2 & 7.4 & 4 & 13.6 & 6.1 & 7.3 \\
    \texttt{ID4} & 40.5 & 10.8 & \SI{6}{\second} & \SI{6}{\second} & 18.3 & 10 & 4 & 10 & 10.3 & 6.3 & 3.9 & 6.4 & 3 & 13 & 5 & 6.5 \\
    \texttt{ID5} & 24.4 & 8.9 & \SI{6}{\second} & \SI{6}{\second} & 19.5 & 11 & 7 & 11.2 & 11.9 & 7.5 & 5.1 & 7.6 & 4 & 14 & 6.3 & 7.5 \\
    \texttt{ID6} & 16.0 & 8.9 & \SI{6}{\second} & \SI{6}{\second} & 19.4 & 10.2 & 7.0 & 11.2 & 9.6 & 8.4 & 5.7 & 8.3 & 4.7 & 14.5 & 5.7 & 7.1 \\
    \texttt{ID7} & 11.7 & 8.9 & \SI{6}{\second} & \SI{6}{\second} & 19.4 & 10.7 & 6.4 & 10.9 & 11.5 & 7.1 & 5.1 & 7.3 & 4 & 13.7 & 5.9 & 7.2 \\
    \texttt{ID7$^\P$}\hspace*{-0.64em} & 11.7 & 8.9 & \SI{6}{\second} & \SI{6}{\second} & 19.5 & 11 & 6.7 & 11 & 12.5 & 7.4 & 5.3 & 7.7 & 4.2 & 13.6 & 6.3 & 7.4 \\
    \texttt{ID8} & 8.3 & 7.5 & \SI{6}{\second} & \SI{6}{\second} &  19 & 10.6 & 6 & 10.7 & 10.9 & 6.8 & 5 & 7.2 & 3.9 & 13.5 & 5.9 & 7.1 \\
    \texttt{ID9} & 8.6 & 7.4 & \SI{6}{\second} & \SI{6}{\second} &  19.1 & 10.1 & 5.1 & 10.2 & 11.1 & 6.6 & 4.5 & 6.6 & 3.7 & 13 & 5.6 & 6.8 \\

    \midrule

    \multicolumn{17}{c}{Continuous Source Separation - MAC reported for \SI{60}{\second} audio sequences} \\
    \texttt{ID1} & 2800 & 43.1 & \SI{6}{\second} & \SI{6}{\second} & {-} & {-} & {-} & {-} & {-} & 7.7 & 5.4 & 7.9 & 4.5 &  14.4 &  6.5 &  7.5 \\
    \texttt{ID1} & 5400 & 43.1 & \SI{6}{\second} & \SI{3}{\second} & {-} & {-} & {-} & {-} & {-} & \bfseries 8 & \bfseries5.7 & \bfseries 8.1 & \bfseries 4.7 &  \bfseries 14.6 &  6.9 & \bfseries 8 \\
    \texttt{ID1} & 10500 & 43.1 & \SI{6}{\second} & \SI{1.5}{\second} & {-} & {-} & {-} & {-} & {-} & \bfseries 8 & \bfseries 5.7 & \bfseries 8.1 & \bfseries 4.7 &  \bfseries 14.6 & \bfseries 7 & \bfseries 8 \\
    \texttt{ID1} & 2700 & 43.1 & \SI{4}{\second} & \SI{4}{\second} & {-} & {-} & {-} & {-} & {-} & 7.5 & 5.2 & 7.6 & 4.2 &  14.2 &  6.4 &  7.3 \\
    \texttt{ID1} & 7500 & 43.1 & \SI{4}{\second} & \SI{2}{\second} & {-} & {-} & {-} & {-} & {-} & 7.8 & 5.5 & 7.9 & 4.5 &  14.6 &  6.8 &  7.7 \\
    \texttt{ID1} & 7700 & 43.1 & \SI{8}{\second} & \SI{8}{\second} & {-} & {-} & {-} & {-} & {-} & 7.7 & 5.5 & 7.9 & 4.6 &  14.4 &  6.3 &  7.3 \\
    \texttt{ID1} & 7600 & 43.1 & \SI{10}{\second} & \SI{10}{\second} & {-} & {-} & {-} & {-} & {-} & 7.8 & 5.4 & 7.9 & 4.5 &  14.3 &  5.9 &  7.1 \\
    \texttt{ID1} & 7300 & 43.1 & \SI{12}{\second} & \SI{12}{\second} & {-} & {-} & {-} & {-} & {-} & 7.6 & 4.5 & 7.8 & 4.5 &  14.2 &  5.6 &  7 \\

    \midrule
    
    \multicolumn{17}{c}{Prompt-Mixture Conditioning Ablation} \\
    \texttt{BLINDPROMPT} & 43.1 & 11.1 & \SI{6}{\second} & \SI{6}{\second} & 19 & 9.7 & 5.4 & 9.7 & 10 & 6.9 & 4.5 & 6.8 & 3.9 & 13.2 &  5.4 &  6.7 \\
    \texttt{INDPROMPT} & 43.1 & 11.1 & \SI{6}{\second} & \SI{6}{\second} & 19.1 & 10.9 & 7.2 & 11.2 & 12 & 7.7 & 5.4 & 7.7 & 4.5 &  14.2 & 6.3 &  7.3 \\
    \texttt{INDALL} & 43.1 & 11.1 & \SI{6}{\second} & \SI{6}{\second} & 18 & 9.9 & 4.3 & 9.6 & 9.4 & 3.4 & 1.5 & 2.3 & 0.7 & 11.6 & 3.5 & 4.3 \\
    \texttt{CAUSAL} & 43.1 & 11.1 & \SI{6}{\second} & \SI{6}{\second} & 18.6 & 10.1 & 5.3 & 10 & 8.6 & 6.4 & 4 & 5.9 & 3.6 & 12 & 4.5 & 5.5 \\

    \midrule
    
    \multicolumn{17}{c}{Simplified Sequence Modeling} \\
    \texttt{ID7 GRU} & 7.2 & 8.8 & \SI{10}{\second} & \SI{10}{\second} & 19.1 & 10.7 & 4.2 & 9.3 & 9 & 6.2 & 4.1 & 6.2 & 3.5 & 11.7 & 4.8 & 5.9 \\
    \texttt{ID7 Hybrid} & 9.2 & 9.2 & \SI{12}{\second} & \SI{12}{\second} & 19 & 10.8 & 6.3 & 10.7 & 11.1 & 6.9 & 4.7 & 7.2 & 4 & 13.6 & 6 & 7.2 \\

    \bottomrule
\end{tabular}
}
\end{table*}

\end{document}